\documentclass{cimento}
\usepackage{graphicx}  

\title{Historical solar Ca II K observations at the Rome and Catania observatories}
\author{T.~Chatzistergos$(^{1,2})$\ETC,
I.~Ermolli\from{ins:x},
M.~Falco\from{ins:a},
F.~Giorgi\from{ins:x},
S.~L.~Guglielmino\from{ins:s},
N.~A.~Krivova\from{ins:y},
P.~Romano\from{ins:a}, \atque
S.~K.~Solanki$(^{2,5})$} 
\instlist{\inst{ins:x} INAF Osservatorio Astronomico di Roma, Monte Porzio Catone, Italy
  \inst{ins:y} Max Planck Institute for Solar System Research,  G\"{o}ttingen, Germany
  \inst{ins:a} INAF Osservatorio Astrofisico di Catania, Catania, Italy
  \inst{ins:s}   Dipartimento di Fisica e Astronomia, Universita degli Studi di Catania, Catania, Italy
  \inst{ins:z} School of Space Research, Kyung Hee University, Yongin, Republic of Korea}

\begin{document}

\maketitle

\begin{abstract}
Here we present the little explored Ca II K archives from the Rome and the Catania observatories and analyse the digitised images from these archives to derive plage areas. 
\end{abstract}

\section{Introduction}
The full-disc Ca II K observations are unique datasets for studies of solar activity and solar irradiance, and hence for understanding the solar influence on Earth's climate.  This is because Ca II K images carry information about the solar magnetic regions and there is a long tradition of observation at this wavelength covering the entire 20th century \cite{chatzistergos_analysis_2017}. 
Various Ca II K archives have recently been digitised, with those from the Kodaikanal, Mt Wilson, Arcetri, and Mitaka observatories being among the most prominent ones.
This led to various studies of the evolution of plage areas, and a number of discrepancies between the published results were noted by \cite{ermolli_comparison_2009,ermolli_potential_2018,chatzistergos_analysis_2018}, most likely due to differences and shortcomings of the processing techniques used by different studies.
Therefore, more data are needed to identify the possible issues with individual data or archives and to fill gaps within the available datasets.

Beside these archives, there are some less explored series such as those from the Rome and Catania observatories, which we introduce here. 
We process the images from these two archives, show results for plage areas and compare them to those derived from the Arcetri data.

\begin{figure}
	\centering
	\includegraphics[width=1\linewidth]{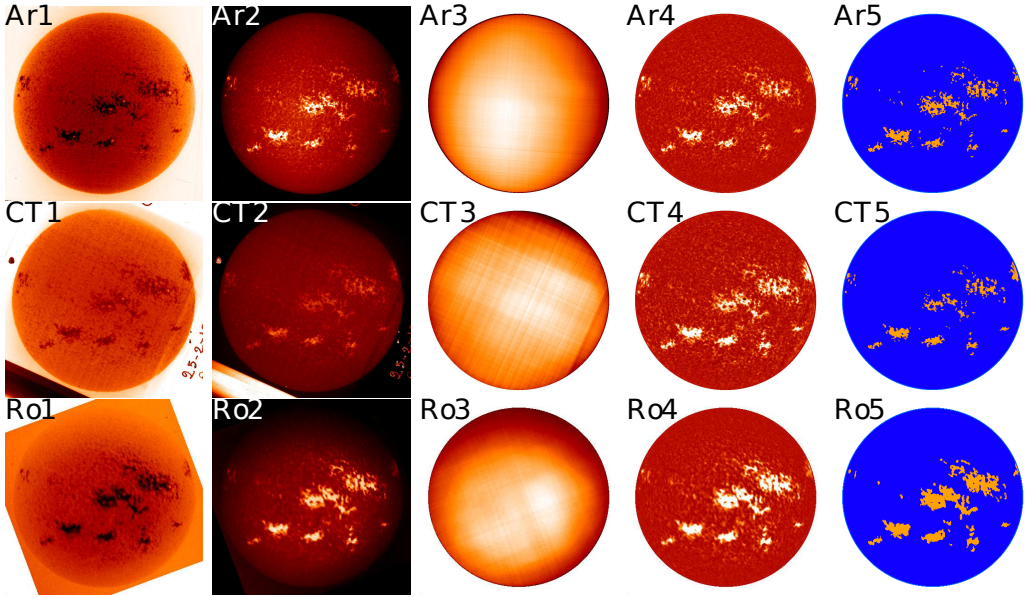}
	\caption{Processing steps on exemplary observation taken on 25/02/1970 from the Arcetri (top), Catania (middle), and Rome (bottom) archives. The columns show the raw negative, raw density image (plotted over the entire density range of the solar disc), identified inhomogeneities including the limb-darkening, photometrically calibrated and limb darkening corrected image (saturated at contrast values [-0.5,0.5]), and segmentation mask (blue is for QS and orange for plage).}
	\label{fig:processingsteps} 
\end{figure}

\section{Data and methods}

The observatory of Rome (Ro, hereafter) at the Monte Mario, in down-town Rome, performed observations in the Ca II K line over the period 1964--1979 with the equatorial spar equipped with a 12 cm objective  (1.8 m focal length).
The observations were made with a Halle birefringent filter with a bandwidth of 0.03 nm \cite{cimino_solar_1967}.
These are high-cadence observations with about 100 images per day.
The photographs were taken with a Cineregistration Northridge camera \cite{cimino_solar_1967} and were recorded in rolls of 35 mm film that are 30 m long each. Each roll contains observations from roughly 8 days.
The diameter of the solar disc on the film is $\sim$16 mm. A sample of these observations were digitised with a commercial 8bit scanner in the year 2000. 
The resulting images have dimensions of $417\times417$ pixel$^2$ and were stored in GIF files.
Examples of the raw images are shown in Fig. \ref{fig:processingsteps}.
The solar disc has a mean radius of 193 pixels, resulting in spatial sampling of $\sim5''$/pixel. 
Only 5823 Ca II K images were scanned, including 1 to 2 observations per day thus covering 2983 days in total. 

The Catania (CT, hereafter) observatory performed observations in the Ca II K line with a spectroheliograph over the period 1907--1977 \cite{zuccarello_solar_2011}. The spectroheliograms were recorded on photographic plates with dimensions $6\times11$ cm$^2$. Only a sample of 92 observations taken between 1970 and 1971 have been digitised with 8 bit accuracy. 
The resulting images have dimensions between $3287\times2102$ and $1972\times1261$ pixel$^2$ and were stored in JPG files.
The radius of the disc varies between 192 and 643 pixels, resulting in a varying spatial sampling between 1.5 and 5''/pixel. 

Here, for comparison purposes we also use the Ca II K data from the Arcetri (Ar, hereafter) observatory. 
These are observations taken with a spectroheliograph and recorded on photographic plates. 
They were taken with a bandwidth of 0.03 nm and have spatial sampling of 2.4$''$/pixel.
There are 5133 observations made over the period 1931--1974, however here we use only the images for the period 1964--1974, i.e. those in the period overlapping with the digitised Ro and CT time series.
More information about the data and the methods used to digitise the Ar data can be found in \cite{ermolli_digitized_2009}.

\begin{figure}
	\centering
	\includegraphics[width=1\linewidth,trim={0 0.0cm 0 0},clip]{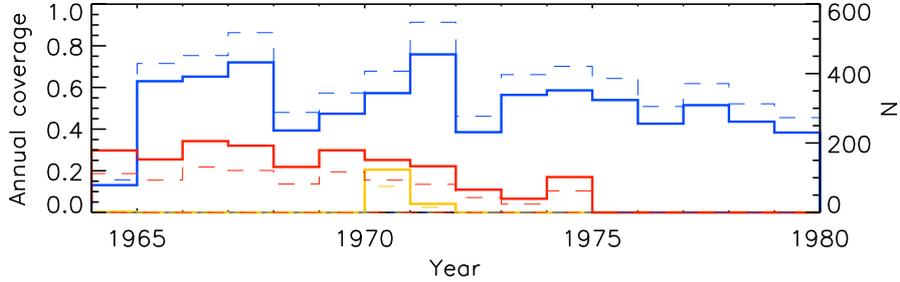}
	\caption{Fraction of days with observations (solid lines and left-hand ordinate) and number of images (dashed lines and right-hand ordinate) for each year in the digitised archives from the Arcetri (red), Catania (yellow), and Rome (blue) observatories.}
	\label{fig:ndata}
\end{figure}

Following the processing applied to Ar data by \cite{chatzistergos_analysis_2018,chatzistergos_analysis_2018-1,chatzistergos_caiik_2018}, we processed the Ro and CT data in exactly the same way. 
Briefly, after converting the images to FITS files, the radius and centre coordinates of the solar disc were identified with a process applying Sobel filtering. The photometric calibration was performed by making use of information stored in the quiet Sun regions. The limb darkening was compensated by dividing the image with a map constructed by iteratively applying a running window filter and fitting 5th order polynomials on horizontal, vertical, and radial directions after the active regions were removed. The identification of the plage regions was done with a multiplicative factor to the standard deviation of the quiet Sun regions. 
Examples of the raw and processed images from the 3 archives are shown in Fig. \ref{fig:processingsteps}. Figure \ref{fig:ndata} shows the number of images per year and the fraction of days in each year with an observation from the 3 archives.
It is evident that the Ro archive provides a better sampling than the Ar and CT ones over that period.

\section{Results}
\begin{figure}
	\centering
	\includegraphics[width=1\linewidth,trim={0 0.1cm 0 0},clip]{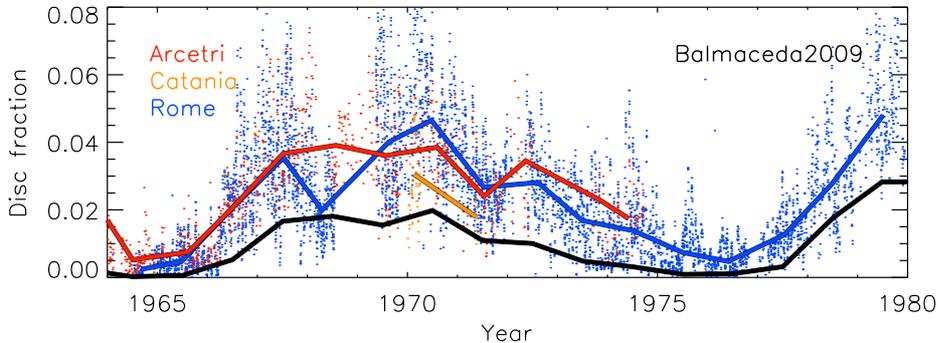}
	\caption{Plage area time-series in fraction of the solar disc from the Arcetri (red), Catania (orange), and Rome (blue) Ca II K data. Daily values are shown in dots, while the solid lines are annual median values. Also shown are the annual values of the sunspot areas by \cite{balmaceda_homogeneous_2009} (black) multiplied by a factor of 10 to put them in the same scale as the plage areas.} 
	\label{fig:1discfractionplage}
\end{figure}

Following \cite{ermolli_comparison_2009}, we computed the power spectrum of the observations within $64\times64$ sub-arrays at the centre of the disc to calculate the frequency which includes 98\% of the power. We found a spatial resolution of $6.7\pm2.3$'', $5.4\pm2.5$'', and $12.8\pm3.4$'',  for the Ar, CT, and Ro data, respectively.
We found images with saturated plage regions in all three archives introduced during the digitisation process.  Since the digitisation of the archives was performed independently and this issue has been reported for other series as well \cite{chatzistergos_analysis_2018}, this appears to be a common problem when digitising Ca II K data.
The limb-darkening compensated images in Fig. \ref{fig:processingsteps} exhibit a lot of similarities, however we note that the sunspots are well seen in the Ar and CT images while they are barely seen in the Ro image. Furthermore, the bright features in the Ro data are more extended than in the Ar or CT ones, which might be due to the different spatial resolution or differences in the bandpass when using a filter instead of spectroheliograph.
Figure \ref{fig:1discfractionplage} shows the derived plage areas in fraction of solar disc as a function of time. We find a good agreement between the Ar and Ro series. We notice that the Ro data give lower plage areas over 1968 but higher over 1970. However, there is a gap in the Ro data between 1968 and 1969, hence the annual values are not necessarily representative. The plage areas from CT are consistently lower than those from Ar or Ro. By keeping only the common days within the Ro and Ar series we find an RMS difference in the plage areas of 0.01, while the mean difference is 0.001 in disc fraction. Comparing the plage areas over the common days in the  CT and Ar series gives RMS difference of 0.017 and mean difference of -0.01 in disc fraction. 

\section{Summary}
We have processed the digitised Ca II K images from the Rome and Catania archives and then derived plage areas. The results show a good agreement with those from the Arcetri observatory. We show that, despite the lower resolution of the Ro data compared to other digitised series,  it is possible to derive accurate results for the plage areas. These archives can be important to fill gaps within the other available Ca II K archives over that period. Furthermore, due to the high cadence of the Ro data, they can potentially be used to study transient phenomena in the solar chromosphere. Digitisation of all available observations from the Rome and Catania observatories would be very valuable.



\begin{thebibliography}{1}
	\expandafter\ifx\csname url\endcsname\relax\def\url#1{\texttt{#1}}\fi
	\expandafter\ifx\csname urlprefix\endcsname\relax\def\urlprefix{URL }\fi
	
	\bibitem{chatzistergos_analysis_2017}
	\NAME{Chatzistergos T.}, {PhD} thesis, 978-3-944072-55-5 (Uni-edition) (2017)
	
	\bibitem{ermolli_comparison_2009}
	\NAME{Ermolli I., Solanki S.~K., Tlatov A.~G., et al.}, \IN{ApJ}{698}{2009}{1000}
	
	\bibitem{ermolli_potential_2018}
	\NAME{Ermolli I., Chatzistergos T., Krivova N.~A. et al.}, 
	Proc. IAU 13 S340 (2018) 115
	
	\bibitem{chatzistergos_analysis_2018}
	\NAME{Chatzistergos T., Ermolli I., et al.},
	\IN{A\&A}{submitted}{AA/2018/34402}
	
	
	\bibitem{cimino_solar_1967}
	\NAME{Cimino M.}, \IN{Solar Physics}{2}{1967}{375}
	
	\bibitem{zuccarello_solar_2011}
	\NAME{Zuccarello F., Contarino L. \atque Romano P.}, \IN{CAOSP}{41}{2011}{85}
	
	\bibitem{ermolli_digitized_2009}
	\NAME{Ermolli I., Marchei E., Centrone M., et al.}, \IN{A\&A}{499}{2009}{627}
		
		\bibitem{chatzistergos_analysis_2018-1}
		\NAME{Chatzistergos T., Ermolli I., Solanki S.~K. et al.},
		\IN{A\&A}{609}{2018}{A92}
		
			\bibitem{chatzistergos_caiik_2018}
		\NAME{Chatzistergos T., Ermolli I., Krivova N.~A. et al.}, 
		Proc. IAU 13 S340 (2018) 125
		
		
	\bibitem{balmaceda_homogeneous_2009}
\NAME{Balmaceda L.~A., Solanki S.~K., Krivova N.~A. et al.},
\IN{JGR}{114}{2009}

	\end{thebibliography}
\end{document}